\documentclass[journal]{IEEEtran}

\usepackage{times}
\usepackage{epsfig}
\usepackage{graphicx}
\usepackage{float}
\usepackage{amsmath}
\usepackage{amssymb,amsfonts,cite}
\usepackage{multirow}
\usepackage[ruled,linesnumbered]{algorithm2e}
\newcommand{\hrulealg}[0]{\vspace{1mm} \hrule \vspace{1mm}}
\usepackage{subcaption}
\DeclareMathOperator*{\argmin}{arg\,min}
\usepackage[breaklinks=true,bookmarks=false]{hyperref}
\usepackage{enumerate}
\usepackage[textwidth=30mm]{todonotes}

\begin{document}

\title{Universal Adversarial Attacks on Neural Networks for Power Allocation in a Massive MIMO System}

\author{Pablo Mill\'{a}n Santos, B.~R.~Manoj,~\IEEEmembership{Member,~IEEE,}  Meysam Sadeghi,~\IEEEmembership{Member,~IEEE,}
       and~Erik~G.~Larsson,~\IEEEmembership{Fellow,~IEEE}
\vspace*{-0.37in}
\thanks{This work was supported in part by Security-Link.}
\thanks{Pablo is with Universidad de Zaragoza, 50009 Zaragoza, Spain and Linköping University,  58183 Link\"{o}ping, Sweden (e-mail: pmillans97@gmail.com).}
\thanks{B. R. Manoj and Erik G. Larsson are with the Department
of Electrical Engineering (ISY), Link\"{o}ping University, 58183 Link\"{o}ping, Sweden (e-mail: \{manoj.banugondi.rajashekara,  erik.g.larsson\}@liu.se).} 
\thanks{Meysam Sadeghi was with Link\"{o}ping University, 58183 Link\"{o}ping, Sweden (e-mail: m.sadeghee@gmail.com).}
}

\maketitle

\begin{abstract}
Deep learning (DL) architectures have been successfully used in many applications including wireless systems. However, they have been shown to be susceptible to adversarial attacks. We analyze DL-based models for a regression problem in the context of downlink power allocation in massive multiple-input-multiple-output systems and propose universal adversarial perturbation (UAP)-crafting methods as white-box and black-box attacks. We benchmark the UAP performance of white-box and black-box attacks for the considered application and show that the adversarial success rate can achieve up to 60\% and 40\%, respectively. The proposed UAP-based attacks make a more practical and realistic approach as compared to  classical white-box attacks.
\end{abstract}


\IEEEpeerreviewmaketitle
\vspace*{-1.3em}
\section{Introduction}
Deep learning (DL) techniques have been widely used in computer vision and natural language processing applications.
Recently, DL-based methods have also exhibited a great success in wireless networks, thanks to their capacity to learn the characteristics of  diverse and complex scenarios. Some of the applications in wireless systems include:  radio-signal modulation classification \cite{bahramali2021robust,sadeghi2018adversarial}, channel estimation and signal detection \cite{bahramali2021robust}, resource allocation \cite{sanguinetti2019deep}, and many more. However, there are robustness concerns around these DL-based techniques  \cite{szegedy2013intriguing}, implying that they are susceptible to the so-called adversarial attacks (well crafted perturbations that cause deep neural network (DNN) based models to deliver erroneous outputs).

Adversarial attacks can be grouped into white-box and black-box attacks, depending on the knowledge of the attacker. In white-box attacks, the attacker knows   the  architecture and parameters. The attacker is usually aware of the inputs of the model, therefore having the full capacity to craft the most harmful attack. On the other hand, in black-box attacks, the victim model parameters are unknown, hence only general input-agnostic attacks can be created. One such attack is universal adversarial perturbation (UAP) which in general is a far more practical and realistic attack as compared to other   white-box attacks found in the literature such as the fast gradient sign method (FGSM), projected gradient descent (PGD), and momentum iterative FGSM (MI-FGSM).  Specifically, the  UAP does not need to know the input samples of the DNN to craft perturbations. This kind of attack can generalize the properties of a small random subset of input samples and generate a universal attack, which in the best scenario (from the attacker's perspective), will fool the model for any input. Furthermore, to craft the UAP black-box attack, since the model is unknown to the attacker, a surrogate model which emulates the original model has to be created. 

Given this security issue of DNNs, the study of adversarial attacks for wireless systems has received considerable attention \cite{bahramali2021robust, sadeghi2018adversarial, manoj2021adversarial}.
Interestingly, almost all   existing work has focused on modulation classification, which is a specific classification task \cite{bahramali2021robust, sadeghi2018adversarial}. However, many problems of interest in wireless systems are regression problems, e.g., resource allocation.

Recently, \cite{manoj2021adversarial}   investigated adversarial attacks against DL-based power allocation algorithms for massive multiple-input-multiple-output (maMIMO) systems. While \cite{manoj2021adversarial} shows that the scope of adversarial attacks is not limited to classification tasks and can cover any regression-type problem, its system model relies on assumptions that limit its practicality. For example, the algorithms proposed in \cite{manoj2021adversarial} require the user positions to be known to the adversary. Further details are given in Section III.B.

In this paper, for a regression problem in  the  context  of  power  allocation  in  a maMIMO  network, we propose two algorithms (one based on accumulative perturbation and the other based on principal component analysis (PCA)) for crafting universal adversarial perturbations (UAPs) in a black-box manner. This means that the adversary neither needs to know the users' positions (input-agnostic attack), nor the DNN model and its parameters (model-agnostic attack). We show that the proposed algorithms can significantly reduce the performance of the system, while they are computationally efficient. Furthermore, we provide a detailed comparison of the performance of UAP attacks with white-box attacks, including FGSM and PGD \cite{manoj2021adversarial}. We also extend   existing white-box attacks, such as the minimum perturbation algorithm \cite{sadeghi2018adversarial} and the optimized attack in \cite{autonomous}, to meet our regression setting.

\vspace*{-1.1em}
\section{System Model}
We consider the downlink of a   multicell maMIMO system using maximum-ratio (MR) and minimum mean-square error (M-MMSE) precoding. The objective of the power control  is to allocate   optimal power to the users by  maximizing  the product  signal-to-interference-plus-noise ratio (SINR), based on knowledge of the positions of the user equipments (UEs). We consider DL-based approach for this task \cite{sanguinetti2019deep}. 

The downlink SINR for a UE $k$ in cell $j$ is given by \cite{sanguinetti2019deep} 
\vspace{-2pt}
\begin{equation} \label{eq:SINR}
\gamma_{jk} = \frac{\rho_{jk}a_{jk}}{\sum^{L}_{l=1}{\sum^{K}_{i=1}{\rho_{li}b_{lijk}+\sigma^{2}}}} \, ,
\vspace{-2pt}
\end{equation}
where $\rho_{jk}$ is the transmission power, $a_{jk}$ is the average channel gain, $b_{lijk}$ is the average interference gain \cite{manoj2021adversarial,sanguinetti2019deep}, $L$ is the number of cells, each cell consists of a base station with $M$ antennas and $K$ number of UEs, and $\sigma^2$ is the noise variance. The max-prod SINR power allocation is defined by 
\vspace{-2pt}
\begin{equation}\label{eq:max_prod}
\hspace{-1em}\max_{\rho_{jk}: \forall{j,k}} \prod^{L}_{j=1}\prod^{K}_{k=1} \gamma_{jk}\, , \quad \text{s.t.} \,\, \sum^{K}_{k=1}{\rho_{jk}} \leq P_\mathrm{max}\,\,, j \in 1,\ldots,L\,,
\vspace{-2pt}
\end{equation}
where $P_\mathrm{max}$ is the maximum downlink transmission power.
\paragraph{DNN architecture}
The DNN model denoted as $f(\cdot,\boldsymbol{\theta})$ is able to learn how to allocate the optimal power for each UE in the cell by only knowing its position, where ${\boldsymbol{\theta}}$ denote the parameters of $f$. To this end, two DNN models have been utilized as shown in Table \ref{tab:DNN}, where the architecture is a feedforward network with all the layers being fully-connected \cite{sanguinetti2019deep}. For brevity, we denote the DNN models $1$ and $2$ as $M_1$ and $M_2$, respectively. 
\begin{table}[t!]
 \begin{center}
  \vspace{-0.1em}
    \caption{Architecture of DNN models.}
    \label{tab:DNN} \scalebox{0.75}{
    \begin{tabular}{|c|c|c|c|} 
    \cline{2-3}
    \multicolumn{1}{c|}{}& Model 1 & Model 2 & \multicolumn{1}{|c}{}\\
    \cline{2-4}
    \multicolumn{1}{c|}{}  & Size & Size & Activation function \\
      \hline
      Input & 40 & 40 & - \\  \hline
      Layer 1 (Dense) & 64 & 512 & elu \\ \hline
      Layer 2 (Dense) & 32 & 256 & elu \\ \hline
      Layer 3 (Dense) & 32 & 128 & elu \\ \hline
      Layer 4 (Dense) & 32 & 128 & elu \\ \hline
      Layer 5 (Dense) & 5  & 5  & elu \\ \hline
      Layer 6 (Dense) & 6  & 6   & linear \\ \hline
    \end{tabular} 
    }
  \end{center}
  \vspace*{-3em}
\end{table}
The input to the DNN model is the UE positions denoted as ${\bf{x}}\in \mathbb{R}^{2KL}$ and the predicted outputs of the model are given by $\boldsymbol{\hat{\rho}}_{j} = [{\hat{\rho}}_{j1}, ..., {\hat{\rho}}_{jK}]$, where $j = 1, \ldots, L$. 
\vspace*{-1.2em}
\section{Adversarial attacks}
In this section, we define the regression-based DNN model as 
$f({\bf{x}}): {\bf{x}} \in {\cal{X}} \rightarrow {\bf{y}} \in {\cal{Y}}$, where 
${\bf{y}}$ is the output of the DNN for the input ${\bf{x}}$,  ${\cal{X}} \in \mathbb{R}^{d}$ is the input domain having dimension $d$, and ${\cal{Y}} \in \mathbb{R}^{c}$ is the output domain with $c$ being the dimension. The idea behind adversarial attacks is to craft an adversarial perturbation $\boldsymbol{\eta}$, which, when added to the input of the DNN yields a wrong output, under the given constraint $||\boldsymbol{\eta}||_{\infty} \leq \epsilon$ -- where $||\boldsymbol{\eta}||_{\infty}$ is the $L_\infty$-norm of the adversarial perturbation and $\epsilon$ is the perturbation magnitude. The so-obtained adversarial  samples can be written as ${\bf{x}}_\mathrm{adv} = {\bf{x}} + {\boldsymbol{\eta}}$, such that $||{\boldsymbol{\eta}}||_{\infty} \leq \epsilon$. 

The aim of crafting adversarial sample is to cause an incorrect  prediction by the  DNN, specifically the  wrong output as $f({\bf{x}}_\mathrm{adv})= {\bf{y}}_\mathrm{adv}$, which is different from that of a prediction on a clean sample $f({\bf{x}}) = {\bf{y}}$. For classification problems, the attack is said to be successful when the output label predicted by $f({\bf{x}}_\mathrm{adv})$ for an adversarial input is different from the output label predicted by $f({\bf{x}})$ for the same input without adversarial perturbations. In contrast, for  a regression  application, many criteria can be chosen to fool the DNN. For our specific application, the attack is said to be successful when the sum of UE powers exceeds $P_\mathrm{max}$, i.e.,  $\sum^{K}_{k=1}{{\hat{\rho}}_{jk}} > P_\mathrm{max}$, which we refer to as infeasible solution. 
  
\vspace*{-1.25em}
\subsection{Universal adversarial perturbation (UAP)}
UAPs were first proposed in  
\cite{moosavi2017universal} for an image classification problem. UAP belongs to a group of attacks characterized by its high capacity to generalize across DNNs, therefore they are specially adequate for black-box attacks. 
In \cite{moosavi2017universal}, the idea is to craft
a universal perturbation progressively by adding a minimum perturbation that moves the adversarial sample to the decision boundaries and after each sum, project it  onto an $\epsilon$-neighborhood. This procedure is repeated until the network gets fooled. For the classification task, the minimum perturbation is generally computed using the DeepFool attack method \cite{deepfool}, whereas for our regression setting, we propose a solution that  extends the method   in \cite{sadeghi2018adversarial}. Furthermore, to craft UAP attacks using the minimum perturbation method through the  algorithm in \cite{deepfool} can be computationally expensive. 
In this regard, 
a simple PCA-based method for a radio-signal classification application was proposed in \cite{sadeghi2018adversarial}. In our paper, we extend this PCA-based method to our specific application with the objective to craft UAPs in a more efficient manner in terms of computational complexity and time. 

Specifically, amongst many, the accumulative perturbation and PCA-based techniques are considered to potentially create universal attacks that have been developed for only the classification tasks, whereas in this paper, we devise implementation methods in the context of a regression-based resource allocation problem, i.e., downlink power allocation in maMIMO. 

\vspace*{-1.25em}
\subsection{Motivation}
For crafting white-box attacks in \cite{manoj2021adversarial}, it is assumed that the UE locations are already available to the adversary. 
In a cellular network, the positions of the UEs can be obtained by sniffing through the physical layer \cite{location_attacks}, that is, listening over the air and decoding the broadcast channel communications (which is typically not  encrypted). There are three different attack modes to obtain the positions \cite{location_attacks}: passive, semi-passive, and active. The passive attack mode consists of decoding the sniffed signals that contain some UE identifiers which are only sent in some areas of the network. With this data, a coarse estimation of the location of the UE can be performed. The semi-passive attack mode improves the precision by taking advantage of some trusted social apps to force the incoming messages from the network to the UEs that are sent in a specific small cell only. Therefore, the UE can be localized with cell precision. Nevertheless, these two attack modes 
may not be useful to craft perturbations in a white-box manner as higher precision (location within a cell) is needed.
With the active attack mode, high precision can be achieved by impersonating the main elements of the network, and, thanks to that asking UEs to send reports which contain positions or at least the signal strength, which can then be used to locate UEs precisely by trilateration. However, conducting this solution to get the UEs' locations does not seem to be reasonable to use in an adversarial attack approach given its complexity. Thus, to overcome this, in this paper, we develop universal adversarial attacks which do not require the UEs positions. 

\vspace*{-1.25em}
\subsection{Attack description}
In Algorithm \ref{alg:Minimum perturbation}, by utilizing \cite{sadeghi2018adversarial}, 
we first provide a method to obtain the minimum perturbation $\boldsymbol{\delta}$. This is further used to craft UAP attacks, that we call accumulative perturbation, given by Algorithm \ref{alg:UAP Moosavi}. We define the loss function for the considered problem as $\mathcal{L}_{j}(f({\bf{x}})) = \sum^{K}_{k=1}{{\hat{\rho}}_{jk}}$, $j = 1, \ldots, L$. The goal of the adversary is to maximize $\mathcal{L}_j(\cdot)$ to achieve an infeasible power output of the DNN model, i.e., for which the sum of the UE powers exceed $P_\mathrm{max}$. In Algorithm \ref{alg:Minimum perturbation}, $\Delta_\mathrm{max}$ is the maximum allowed perturbation norm, which is the maximum magnitude that the perturbation can have in the infinity norm sense and $\epsilon_\mathrm{acc}$ is the desired perturbation accuracy which is initialized to a very small value. The maximum number of iterations to compute the minimum perturbation is $\mathrm{I_{max}}$. In the algorithm, we have employed the $L_\infty$-norm, thus utilizing the FGSM, i.e., the sign of the gradient of ${\mathcal{L}}_j(\cdot)$.
The process is repeated until $\epsilon_\mathrm{acc}$ is fulfilled or $\mathrm{I_{max}}$ is reached. The output of the algorithm is the
FGSM perturbation resulting in the minimum perturbation. 
\begin{algorithm}[t]
\small 
\caption{Minimum perturbation algorithm min\_perturbation$(f, {\bf{x}}, \epsilon, P_\mathrm{max})$}
    \label{alg:Minimum perturbation}
    \SetAlgoNoLine
    \SetKwInOut{Input}{Input} 
    \SetKwInOut{Output}{Output}
    \SetKwRepeat{Do}{do}{while}
    \Input{${\bf{x}}$, $P_\mathrm{max}$, $f(\cdot,{\bf{\theta}})$, $\Delta_\mathrm{max}$, $\epsilon_\mathrm{acc}$, $\mathrm{I_{max}}$ }
    \Output{$\bf{\delta}$}
    \hrulealg
    \text{\textbf{Initialize}: $\epsilon_\mathrm{max} = \Delta_\mathrm{max}$, $\epsilon_\mathrm{min} = 0$, $\mathrm{I} = 0$ }\\
    \While {$(\epsilon_\mathrm{max} - \epsilon_\mathrm{min} > \epsilon_\mathrm{acc})$ {\bf{and}} $(\mathrm{I} < \mathrm{I_{max}})$}{
    $\mathrm{I} = \mathrm{I} + 1$ \\
    $\epsilon = (\epsilon_\mathrm{max} + \epsilon_\mathrm{min})/2$ \\
    ${\bf{x}}_\mathrm{adv} = {\bf{x}} + \epsilon \cdot \mathrm{sign}(\nabla_{{\bf{x}}}{\mathcal{L}_j(f({\bf{x}}))})$ \\
    \eIf{$\sum^{K}_{k=1}{{\hat{\rho}}_{jk}} < P_\mathrm{max}$}{
    $\epsilon_\mathrm{min} = \epsilon$}{
    $\epsilon_\mathrm{max} = \epsilon$
    }
    }
    $\boldsymbol{\delta} = \epsilon_\mathrm{max} \cdot \mathrm{sign}(\nabla_{{\bf{x}}}{\mathcal{L}_j(f({\bf{x}}))})$ \\
    \text{\textbf{return} $\boldsymbol{\delta}$}
    \vspace{-2pt}
\end{algorithm}

The method to craft UAP attacks is shown in Algorithm \ref{alg:UAP Moosavi}, where $\{{\bf{x}}_{1},...,{\bf{x}}_{N}\}$ are the $N$ random input samples taken from the training dataset and $\epsilon$ is the desired magnitude of the universal perturbation $\boldsymbol{\eta}$. $\mathrm{I_{max}}$ is the maximum number of iterations and we use the $L_{\infty}$-norm. 
The function $\mathrm{clip}(\cdot)$ is used to perform the clipping operation which projects $\boldsymbol{\eta}$ onto an $L_{\infty}$-norm of $\epsilon$ radius.
The idea here is to find the minimum perturbation $\boldsymbol{\delta}$  for each input sample ${\bf{x}}_\mathrm{t}$, where $t = 1,\ldots, N$ and accumulate the perturbations in a resultant universal perturbation $\boldsymbol{\eta}$ if the previous value of $\boldsymbol{\eta}$ does not fool the model. Also, each time $\boldsymbol{\delta}$ is accumulated, the result is projected onto the $L_{\infty}$ $\epsilon$-neighborhood.  
The algorithm proceeds for $\mathrm{I_{max}}$ iterations.
\begin{algorithm}[t]
\small 
    \caption{Accumulative perturbation-based approach for crafting UAP}
    \label{alg:UAP Moosavi}
    \SetAlgoNoLine
    \SetKwInOut{Input}{Input} 
    \SetKwInOut{Output}{Output}
    \SetKwRepeat{Do}{do}{while}
    \SetKwFor{For}{for}{do}{endfor}
    \Input{$\{{\bf{x}}_{1},...,{\bf{x}}_{N}\}$, $P_\mathrm{max}$, $f(\cdot,{\bf{\theta}})$, $\epsilon$, $\mathrm{I_{max}}$}
    \Output{${\boldsymbol{\eta}}$}
    \hrulealg
    \text{\textbf{Initialize}: $\boldsymbol{\eta} = {\boldsymbol{0}}_{2KL}$, $\mathrm{I} = 0$ }\\
    \While {$(\mathrm{I} < \mathrm{I_{max}})$}{
    \For{$i$ $\mathrm{in}$ $range(N)$}{
    ${\bf{x}}_\mathrm{adv} = {\bf{x}}_{i} + \boldsymbol{\eta}$ \\
    \uIf{$\sum^{K}_{k=1}{{\hat{\rho}}_{jk}} \leq P_\mathrm{max}$}{
    $\boldsymbol{\delta} = $ min\_perturbation$(f, {\bf{x}}_\mathrm{adv}, \epsilon, P_\mathrm{max})$\\
    $\boldsymbol{\eta} = \boldsymbol{\eta} + \boldsymbol{\delta}$\\
     $\boldsymbol{\eta}= {\mathrm{clip}}(\boldsymbol{\eta}, -\epsilon, \epsilon)$
    }
    } 
    $\mathrm{I} = \mathrm{I} + 1$\\
    }
    \text{{\bf{return}} ${\boldsymbol{\eta}}$}
\end{algorithm}
Building upon  \cite{sadeghi2018adversarial},  we now develop a method called a PCA-based UAP attack, which is more efficient in terms of computational complexity   compared to  Algorithm \ref{alg:UAP Moosavi}. 
In Algorithm \ref{alg:UAP},  $\{{\bf{x}}_{1},...,{\bf{x}}_{N}\}$ are the $N$ randomly chosen input samples from the training dataset ${\bf{x}}$ and $\epsilon$ is the desired perturbation magnitude. The samples $\{{\bf{x}}_{1},...,{\bf{x}}_{N}\}$ are used to find the directions of the perturbations $\{{\bf{n}}_\mathrm{{x_{1}}}, ..., {\bf{n}}_\mathrm{{x_{N}}}\}$, where  ${\bf{n}}_\mathrm{{x_{t}}}$ is the gradient of the loss function, $t = 1, \ldots, N$. This method uses a PCA-based approach, which performs an orthogonal linear transformation that maps the data to a new coordinate system, where the first component has the largest variance, the second component the second largest variance, and so forth. From $\{{\bf{n}}_\mathrm{{x_{1}}}, ..., {\bf{n}}_\mathrm{{x_{N}}}\}$, we construct a matrix,  and apply PCA to select the first principal component. With this, we   obtain a general perturbation direction that represents the common properties of ${\bf{n}}_\mathrm{{x_{t}}}$.
Finally, to craft the universal perturbation, we  employ FGSM since the $L_\infty$-norm is utilized. 

\begin{algorithm}[t!]
\small 
    \caption{PCA-based approach with $L_{\infty}$-norm for crafting UAP}
    \label{alg:UAP}
    \SetAlgoNoLine
    \SetKwInOut{Input}{Input} 
    \SetKwInOut{Output}{Output} 
    \Input{ $\{{\bf{x}}_{1},...,{\bf{x}}_{N}\}$, $P_\mathrm{max}$, $f(\cdot,{\boldsymbol{\theta}})$, $\epsilon$}
    \Output{${\boldsymbol{\eta}}$}
    \hrulealg
    Evaluate ${\bf{X}}^{N \times 2KL} = [{\bf{n}}_\mathrm{{x_{1}}}, ..., {\bf{n}}_\mathrm{{x_{N}}}] ^\mathrm{T}  = [\nabla_{{\bf{x}}_{1}}\mathcal{L}_j(f({\bf{x}}_{1})),..., \nabla_{{\bf{x}}_{N}}\mathcal{L}_j(f({\bf{x}}_{N}))]^\mathrm{T}$\\
    Compute the first principal direction of ${\bf{X}}$ and denote it by ${\bf{v}}_{1}$, i.e., ${\bf{X}  = U \Sigma V}^\mathrm{T}$ and ${\bf{v}}_{1} = {\bf{V}}(:,1)$\\
    $\boldsymbol{\eta} = \epsilon \cdot$ sign$({\bf{v}}_{1})$\\
    \text{\textbf{return} $\boldsymbol{\eta}$}
    \vspace{-2pt}
\end{algorithm}
\vspace*{-1.25em}
\subsection{Countermeasures}
Countermeasures may  be employed to mitigate the adversarial attacks. The most common method of defense is to use adversarial training, which is to generate adversarial samples using a gradient-based method and augment it with the clean samples during the standard training process to obtain a robust DNN \cite{adversarial_training}. Another possible defense is to identify the perturbation and eliminate it from the original data, which is called  perturbation subtraction defense \cite{bahramali2021robust}. Other defense methods  include  random feature pruning, randomization, and denoising  \cite{ren2020adversarial}. We could also consider scaling the predicted powers to avoid infeasibility, which would be an application-specific solution but further investigation is required about its effectiveness.  
A detailed investigation of possible countermeasures to our proposed attacks
is beyond the scope of this paper. 

\vspace*{-1.1em}
\section{Experiments and discussion}
\paragraph{Dataset}
We utilize the openly available dataset for DL-based downlink power allocation in maMIMO and the DNN models of \cite{sanguinetti2019deep} to evaluate the robustness against the attacks that we have developed in this paper.
The specific values of the network are: $L = 4$ cells, where each cell coverage area is $250$ m $\times$ $250$ m and each cell is served by   a base station with $M = 100$ antennas.
Each cell has  $K = 5$ UEs. The communication bandwidth is $20$ MHz, the noise power is $\sigma^{2} = -94$ dBm and the maximum downlink transmission power is $P_\mathrm{max}=500$ mW. The dataset used to train the DNN models has $N_{T} = 329,000$ samples, where each sample contains the UEs' positions $\{{\bf{x}}(n);n = 1,..., N_{T}\}$ as input and the corresponding power allocation  $\{{\boldsymbol{\rho}}_{j}(n)=[\rho_{j1},\ldots,\rho_{jK}] \in \mathbb{R}^K, n = 1,..., N_{T}\,\, \text{and}\,\, j = 1,\ldots,L\}$ obtained using \eqref{eq:max_prod} as output. The solution to \eqref{eq:max_prod} is obtained by solving  classical optimization methods \cite{sanguinetti2019deep}. This means that, $\boldsymbol{\rho}_{j}$'s are the true transmission powers used to train the models.

Each input sample to the DNN is ${\bf{x}}\in \mathbb{R}^{2KL}$ corresponding to the position of all the UEs in a maMIMO network. Each output of the DNN for a given cell $j$ is  ${\boldsymbol{{\rho}}_{j}}\in\mathbb{R}^{K+1}$; $K+1$ is due to the additional information that the sum of all the UEs' powers in each cell being constrained to $P_\mathrm{max}$.
We can write $\boldsymbol{{\rho}}_{j} = [{{\rho}}_{j1}, ..., {{\rho}}_{jK}, \sum^{K}_{k=1}{{{\rho}}_{jk}}]$. The DNN model is able to satisfy the power limit of $P_\mathrm{max}$ and produces better prediction that is comparable to a model-based approach. The dataset has been split into training and test samples, where the training dataset has $329,000$ samples while the test has $500$ samples. 

\paragraph{Results}
In this section, we evaluate the robustness of the model $M_1$ by using the proposed universal adversarial attack methods. We perform both white-box and black-box attacks, and analyze the results for MR and M-MMSE precoding schemes. For the UAP,  $N = 1500$ random  samples are chosen from ${\bf{x}}$.
In the evaluation, we denote the test dataset by ${\bf{x}}_\mathrm{test}$ and take the number of samples in this set to be $500$. These test samples were cross-validated to make sure they provide feasible outputs on models $M_1$ and $M_2$, which is as expected. Moreover, for each cell, the UAP Algorithms \ref{alg:UAP Moosavi} and  \ref{alg:UAP} are performed with Monte-Carlo iterations (e.g., $50$) and we select the best perturbation that gives more infeasible outputs amongst the iterations. For evaluation, we consider different values for the perturbation magnitude: $\epsilon =  \{0.1, 0.2, 0.3, 0.4, 0.5, 0.7, 1\}$.

Firstly, UAP white-box attacks were crafted using both proposed methods (Algorithms \ref{alg:UAP Moosavi} and  \ref{alg:UAP}). The algorithms are applied against $M_1$ to obtain the universal adversarial attacks. The resulting perturbation   is applied to the test samples ${\bf{x}}_\mathrm{test}$ to analyze the infeasible outputs provided by the model, which evaluates the susceptibility of the model in terms of adversarial success rate against UAP. The DNN model ${M}_{2}$ has been used as a surrogate model for applying the black-box attacks. This model has more trainable parameters than ${M}_{1}$, which was mainly used by \cite{sanguinetti2019deep} to obtain a better estimate of the power allocation coefficients. 
By using transferability property \cite{manoj2021adversarial}, we obtain the universal perturbations using the proposed methods from a substitute model $M_2$. Thus, the attack is crafted without any knowledge of the victim model $M_1$. 
Once the UAPs are built, they are evaluated on ${M}_{1}$ as done in the white-box attacks. 

We compared our obtained performance results  with the other white-box attack methods, such as FGSM and PGD proposed in \cite{manoj2021adversarial} for the same application. We further extend the comparison to 
the minimum perturbation-based (Algorithm \ref{alg:Minimum perturbation}) and the optimization-based white box-attacks (Algorithm \ref{alg:optimized}). In Algorithm \ref{alg:optimized}, we extend \cite{autonomous} to our specific regression problem by solving the proposed loss function  ${\mathcal{L}}_j(\cdot)$ using the Adam optimizer.  
\begin{algorithm}[!t]
\small 
    \caption{Optimized white-box attack}
    \label{alg:optimized}
    \SetAlgoNoLine
    \SetKwInOut{Input}{Input} 
    \SetKwInOut{Output}{Output}
    \SetKwFor{For}{for}{do}{endfor}
    \Input{
    ${\bf{x}}$, $P_\mathrm{max}$, $f(\cdot, \boldsymbol{\theta})$, $\epsilon$, $\mathrm{I}_\mathrm{max}$
   }
    \Output{$\boldsymbol{\eta}$}
    \hrulealg
    \textbf{Initialize}: $\boldsymbol{\eta} = {\boldsymbol{0}}_{2KL}$\\
    \For{$i$ $\mathrm{in}$ $range(\mathrm{I_{max})}$}{
    ${\bf{x}}_\mathrm{adv} = {\bf{x}}_{i} + \boldsymbol{\eta}$ \\
    ${\bf{x}}_\mathrm{adv} = \mathrm{clip}({\bf{x}}_\mathrm{adv}, {\bf{x}}_{i} - \epsilon, {\bf{x}}_{i} +\epsilon)$\\
    Obtain $\argmin_{\boldsymbol{\eta}}{-\mathcal{L}_j(f({\bf{x}}_\mathrm{adv}))}$\\
    \uIf{$\sum^{K}_{k=1}{{\hat{\rho}}_{jk}} > P_\mathrm{max}$}{
    \textbf{break}}
    }
    \textbf{return} $\boldsymbol{\eta}$
    \vspace{-2pt}
\end{algorithm}
\begin{figure*}[t]
    \centering
    \begin{subfigure}[b]{0.49\textwidth}
       \centering
        \centerline{\includegraphics[width=3.5in,height=1.8in]{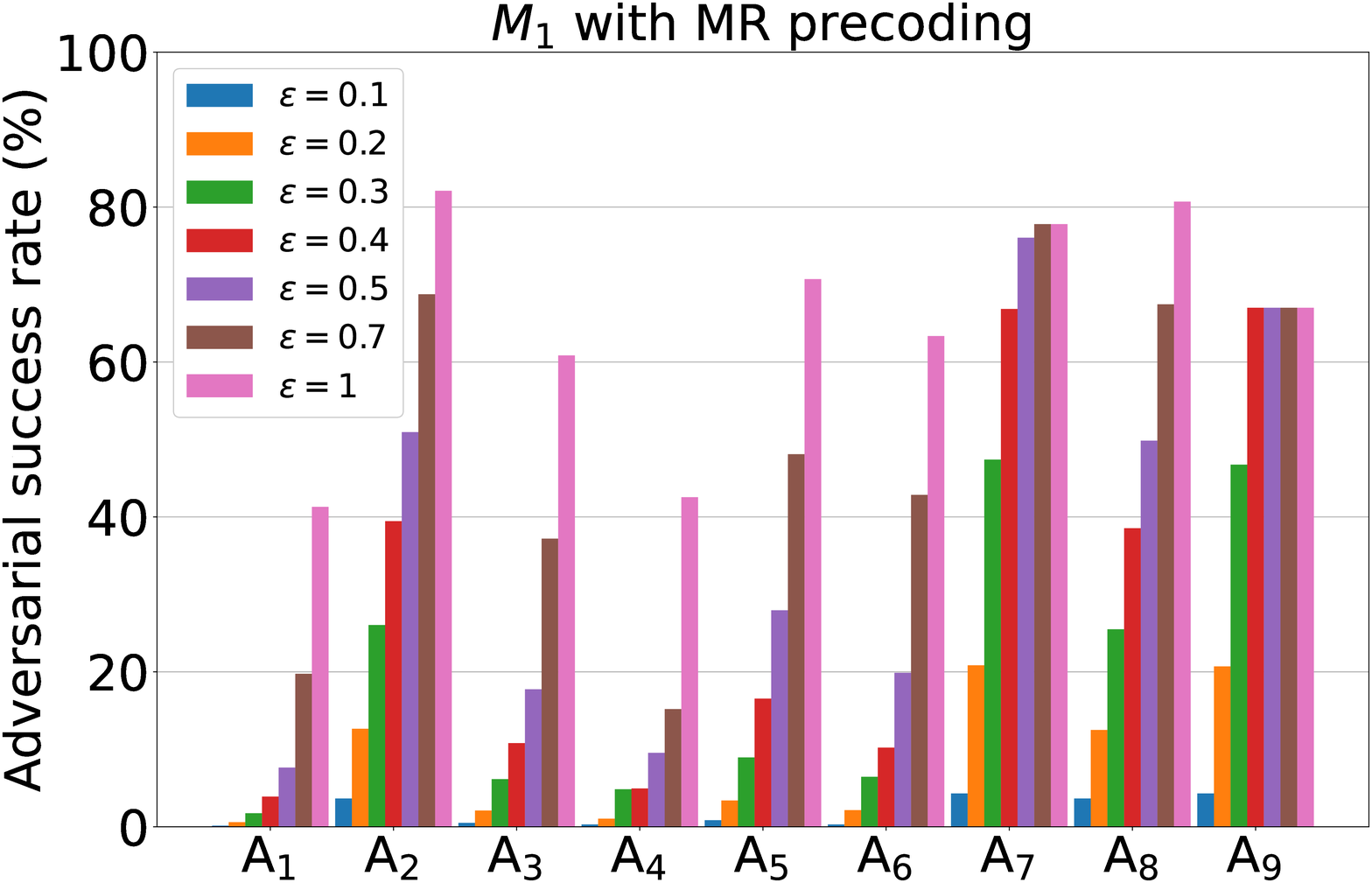}\vspace*{-0.15in}}
        \caption{MR precoding.}
        \vspace*{-0.15in}
        \label{fig:Bar plot all MR}
    \end{subfigure}
    \hfill
    \begin{subfigure}[b]{0.49\textwidth}
        \centering
        \centerline{\includegraphics[width=3.5in,height=1.8in]{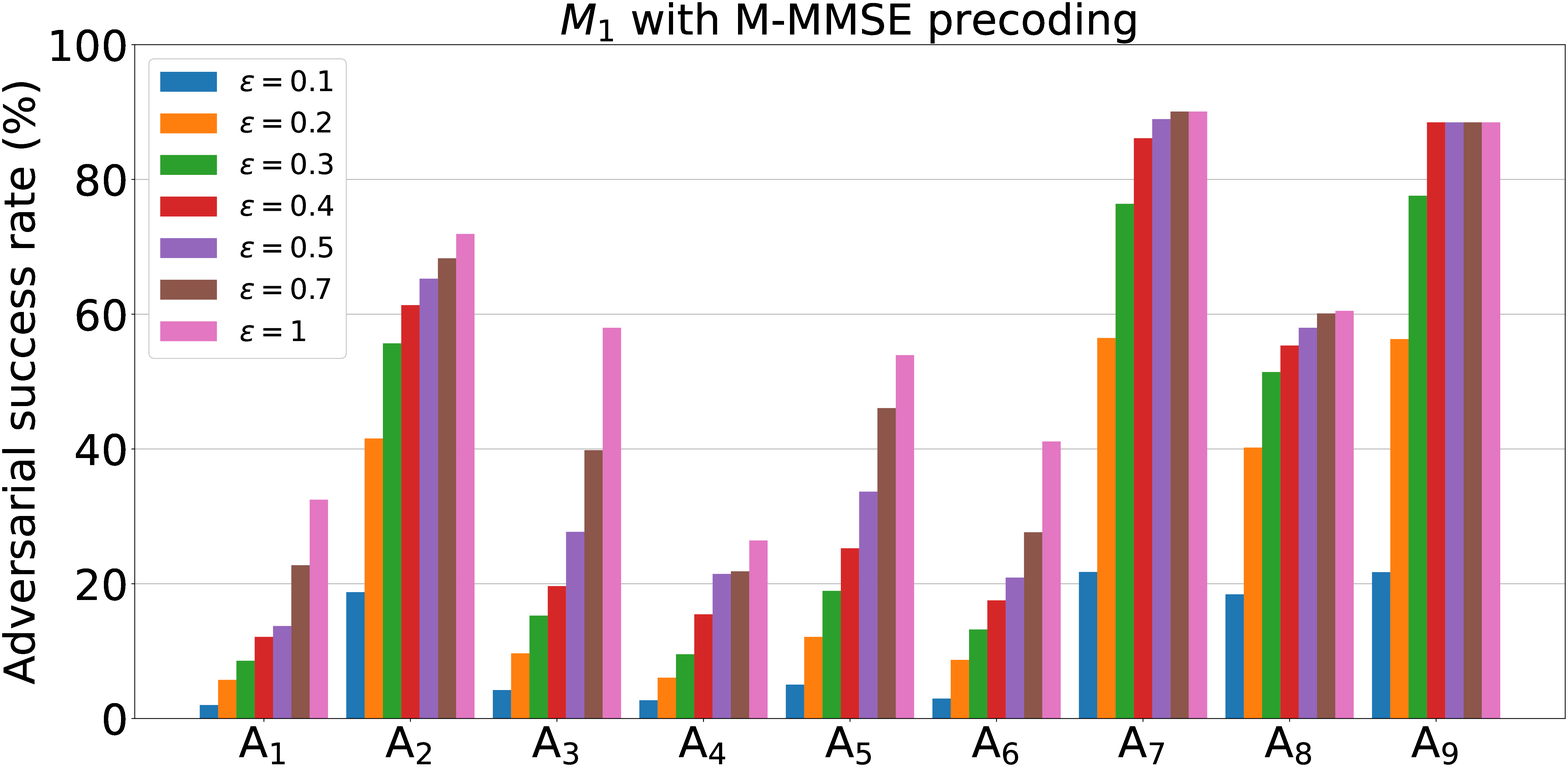}\vspace*{-0.15in}}
        \caption{M-MMSE precoding.}
        \vspace*{-0.15in}
        \label{fig:Bar plot all MMMSE}
    \end{subfigure}
    \vspace*{0.1in}
    \caption{Adversarial success rate of different attack methods.}
    \vspace*{-0.2in}
    \label{fig:attacks}
\end{figure*}
\begin{table}[h!]
 \begin{center}
  \vspace{-0.15em}
    \caption{Notations of attack methods.}
      \vspace{-0.15em}
    \renewcommand{\arraystretch}{1.15}
    \label{tab:attacks_notations} \scalebox{0.8}{
    \begin{tabular}{|c|c|}
    Notation & Attack methods\\
      \hline
      $A_1$ & Random perturbation \\
      \hline
      $A_2$ & Minimum perturbation (Algorithm \ref{alg:Minimum perturbation}) \\
      \hline
      $A_3$ & Accumulative perturbation UAP white-box (Algorithm \ref{alg:UAP Moosavi})\\
      \hline
      $A_4$ & Accumulative perturbation UAP black-box (Algorithm \ref{alg:UAP Moosavi}) \\
      \hline
      $A_5$ & PCA-based UAP white-box (Algorithm \ref{alg:UAP}) \\
      \hline
      $A_6$ & PCA-based UAP black-box (Algorithm \ref{alg:UAP}) \\
      \hline
      $A_7$ & Optimized white-box attack (Algorithm \ref{alg:optimized}) \\
      \hline
      $A_8$ & White-box attack: FGSM \\
      \hline
      $A_9$ & White-box attack: PGD \\
      \hline
    \end{tabular} }
  \end{center}
    \vspace*{-1em}
\end{table}
For convenience, we use the notation   $A_m$, $m = 1,\ldots, 9$ to denote different attack methods; see  Table \ref{tab:attacks_notations}. 
Fig. \ref{fig:Bar plot all MR} presents the adversarial success rate on $M_1$ by using MR precoding and for  the attack methods  given in Table \ref{tab:attacks_notations}. 

As expected, white-box attacks ($A_2,A_7,A_8,A_9$) give higher adversarial success rates than random perturbations ($A_1$) and UAP attacks ($A_3-A_6$).
A minimum perturbation-based attack ($A_2$) computes the minimum $\epsilon$ that fools the DNN. For example, when the attack rate is plotted for $\epsilon = 0.3$,   the largest $\epsilon$ used by   Algorithm \ref{alg:Minimum perturbation} ($A_2$) is $0.3$, but the algorithm could have found even smaller  values of $\epsilon$   to fool the model which is different from $A_8$ that uses the fixed value of $\epsilon$. We noticed that PGD ($A_9$) achieves the highest attack rate. The optimized attack ($A_7$) yields  very similar performance to $A_9$, possibly because when performing the clipping operation, the solution to the minimization problem is similar to the perturbation obtained with $A_9$. The UAP white-box attacks have less adversarial success rate than the white-box attacks, which is as expected and intuitive. To create the UAP attacks, $1500$ random samples were chosen from the training dataset and the so-obtained perturbation was applied on the test samples; however, if the perturbation had been crafted from the test samples itself, then we might get a better adversarial attack rate than with the former approach. The PCA-based UAP method ($A_5$) achieves a better attack rate  compared to the accumulative perturbation-based  method ($A_3$). 

In \cite{sadeghi2018adversarial}, for a classification task, the black-box UAPs show a much smaller attack rate  compared to the white-box attacks, which is intuitively satisfying. In this regard, we further observed that for black-box attacks, the PCA-based ($A_6$) has higher adversarial impact than the accumulative perturbation method ($A_4$), which means that performing PCA allows to generalize better the properties of the input samples  compared to accumulating the minimum perturbation for each input sample. Another observation is that the difference between the UAP white-box ($A_3, A_5$) and black-box attacks ($A_4, A_6$) using Algorithm \ref{alg:UAP Moosavi} is almost the same as for Algorithm \ref{alg:UAP}, implying that both these algorithms have similar transferability properties.
\begin{figure}[t]
 \vspace*{-0.09in}
    \centering
        \centerline{\includegraphics[width=3.7in, height=1.9in]{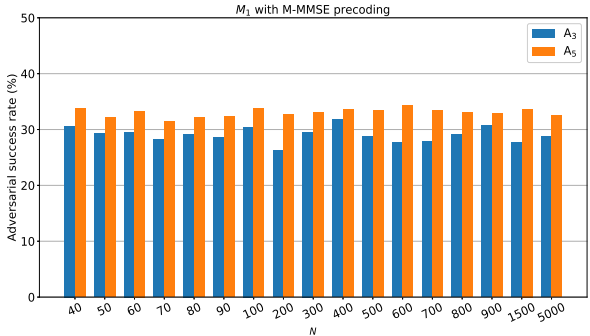}
        \vspace*{-0.08in}
        }
        \caption{Adversarial success rate of UAPs for different $N$.}
        \label{fig:UAP_N}
      \vspace{-0.4em}
\end{figure}

Fig. \ref{fig:Bar plot all MMMSE} presents the adversarial success rate on $M_1$ by using M-MMSE precoding  for different attack methods as shown in Table \ref{tab:attacks_notations}. The M-MMSE precoding scheme  takes interference also into account and is more complex compared to MR. The DL solution for the power allocation is better with M-MMSE than with MR precoding \cite{sanguinetti2019deep}, so the DNN predictions with M-MMSE yield the sum-UE powers much closer to $P_{\mathrm{max}}$. However, this fact also makes the DNN model more vulnerable to adversarial perturbations because the solutions are closer to infeasibility.

The attacks $A_2$ and $A_8$ have a similar performance, since both are FGSM-based technique. Similarly, the white-box attacks $A_7$ and $A_9$ are comparable, providing the highest attack rate amongst all white-box and black-box attacks.
In the UAP black-box attacks scenario, Algorithm \ref{alg:UAP Moosavi} ($A_3$) achieves a lower adversarial rate than Algorithm \ref{alg:UAP} ($A_5$). However, the UAP methods have much higher attack rates than the random perturbations, which confirms the success of well-crafted universal adversarial attacks. 
Furthermore, we have compared the attack rates of $A_{3}$ and $A_{5}$ for different $N$ as shown in Fig. \ref{fig:UAP_N}. The results show a small variability, $\pm 3\%$ in both attacks, demonstrating that universal perturbations can be crafted with a few samples. Moreover, we have estimated the execution time per cell of each algorithm for the assumed parameters used in Fig. \ref{fig:Bar plot all MMMSE}; this is summarized in \autoref{tab:time}. It can be seen that $A_{3}$ is substantially slower than $A_{5}$.
\begin{table}[t!]
 \begin{center}
  \caption{Attack time-complexity per cell.}
  \label{tab:time}
  \vspace{-0.1em}
  \renewcommand{\arraystretch}{1.1}
\label{tab:attacks} \scalebox{0.78}{
        \begin{tabular}{|c|c|c|c|c|c|c|c|}
        \small
        Attack methods &  $A_1$ & $A_2$ & $A_3$ & $A_5$ & $A_7$ & $A_8$ & $A_9$\\
          \hline
          Time in sec & 0.003 & 12.57 &  2886.26 & 358.36 & 131.75 & 2.05 & 95.142 \\
      \hline
        \end{tabular} }
 \end{center}
 \vspace{-2em}
\end{table}

\vspace*{-1.1em}
\section{Conclusion}
\vspace*{-0.1em}
We developed new adversarial attacks for the regression  problem that arises  in the context of power allocation in  maMIMO systems.
Specifically, we  proposed universal attacks which are practical and realizable, that we termed accumulative perturbation-based and PCA-based approaches. Using our proposed UAP algorithms, it is possible to generate   adversarial perturbations even with a few random samples, while being input-agnostic.
Moreover, for both MR and M-MMSE precoding schemes, the UAP white-box and black-box ($M_2$ is used as a surrogate model while $M_1$ is the victim model) attacks were compared with the other non-universal attacks. In addition to these, we   also proposed an optimization-based white-box attack (Algorithm \ref{alg:optimized}). In conclusion, with our proposed attacks, significant success rates (for the attacker) can be achieved, without knowing the UE positions. This exposes a new type of security threat to wireless systems that rely on DNN to solve resource allocation problems.

\vspace*{-0.75em}
\bibliographystyle{IEEEtran}
\vspace{-0.3em}
\bibliography{bibliography}
\end{document}